\documentclass[superscriptaddress,twocolumn,prl,showpacs]{revtex4}
\usepackage[dvips]{graphicx}
\usepackage{amsmath}
\usepackage{bm}
\usepackage{braket}

\def\beq{\begin{equation}}
\def\eeq{\end{equation}}
\def\bea{\begin{eqnarray}}
\def\eea{\end{eqnarray}}
\def\nn{\nonumber\\}
\def\rvec{\bm{r}}
\def\Rvec{\bm{R}}
\def\Avec{\bm{A}}
\def\Bvec{\bm{B}}
\def\Pvec{\bm{P}}
\def\pvec{\bm{p}}
\def\Vvec{\bm{V}}
\def\Fvec{\bm{F}}

\def\chivec{\bm{\chi}}
\def\Omegavec{\bm{\Omega}}

\def\Hcal{\mathcal{H}}
\def\HcalN{\mathcal{H}_\mathrm{N}}
\def\el{\mathrm{el}}
\def\psiel{\psi_\mathrm{el}}
\def\varphitilde{\widetilde{\varphi}}

\begin{document}
\title{Electron corrected Lorentz forces in solids and molecules in
magnetic field}
\author{Davide Ceresoli}
\affiliation{Scuola Internazionale Superiore di Studi Avanzati (SISSA)
and DEMOCRITOS, via Beirut 2-4, 34014 Trieste, Italy}
\author{Riccardo Marchetti}
\affiliation{Scuola Internazionale Superiore di Studi Avanzati (SISSA)
and DEMOCRITOS, via Beirut 2-4, 34014 Trieste, Italy}
\author{Erio Tosatti}
\affiliation{Scuola Internazionale Superiore di Studi Avanzati (SISSA)
and DEMOCRITOS, via Beirut 2-4, 34014 Trieste, Italy}
\affiliation{International Center for Theoretical Physics (ICTP),
Strada costiera 11, 34104 Trieste, Italy}
\date{\today}

\begin{abstract}
We describe the effective Lorentz forces on the ions of a generic
insulating system in an magnetic field, in the context
of Born-Oppenheimer ab-initio molecular dynamics. The force on each ion
includes an important contribution of electronic origin, which depends
explicitly on the velocity of all other ions. It is formulated in terms of
a Berry curvature, in a form directly suitable for future first principles
classical dynamics simulations based {\it e.g.,} on density functional
methods. As a preliminary analytical demonstration we present the dynamics
of an H$_2$ molecule in a field of intermediate strength, approximately
describing the electrons through Slater's variational wavefunction.
\end{abstract}
\maketitle

Understanding the behavior of matter in large magnetic fields is
important both conceptually and practically.  The effect of a magnetic
field on a non-magnetic (molecular, fluid or solid) electronic insulator
is twofold.  The first is on the electronic states, via a field induced
modification of electron quantization. The states of free electrons in
a magnetic field are split into Landau levels; in a general insulator,
the electronic states or bands will turn into some generalized form
of Landau levels, whose splittings are important when at large fields
they become comparable with band gaps.  These electronic effects can be
efficiently calculated with modern density functional (DFT) methods well
described in the literature.~\cite{mauri_cai04}.  The second type of field
effects, important in dynamics, are those on nuclear motion, via Lorentz
forces. In an insulating or close shell system, where electron motion
can be decoupled adiabatically, the effective Lorentz force $\Fvec_n$
on any ion $n$ still depends on the electronic states.  Schematically,
the total effective Lorentz force may range between the bare Lorentz
force $\Fvec_n = (Q_n/c)\Vvec\times\Bvec$ (if electrons could be ignored)
to zero (if electrons could be assumed to be tightly bound to nuclei,
thus neutralizing them).  Even if the effect on the electronic structure
is very small and probably negligible for solid state applications, the
electronic contribution to the Lorentz force is not negligible at all,
but is so far without a clear understanding of how one may, at least in
principle, calculate it.

A pioneering fully quantum-mechanical treatment of nuclei and electrons
in a magnetic field was formulated long ago, establishing a clear basis
of principle.~\cite{schmelcher88-89,schmelcher97}.  The separation of the
center of mass nuclear motion gives rise, apart from a trivial motional
Stark effect, to a number of mass-correction terms in the electronic and
nuclear hamiltonians, which provide additional couplings of the motion
of different nuclei. Unfortunately, this approach is computationally
very demanding and can be applied at best to small molecules or clusters.
There exists to our knowledge neither an explicit formulation that can
be used right away for ab-initio simulations, for example of DFT type,
nor a direct application to real systems, beyond the hydrogen
atom~\cite{resta00}.

In this letter we pursue a formulation of electronic Lorentz forces based
on the Born-Oppenheimer approximation, in principle suitable for
state-of-the-art first principles simulations. As an illustration,
we apply it to the dynamics in a magnetic field of the $H_2$
molecule described by a simple variational wavefunction.  
Following~\cite{resta00,yin94}, we will use the electronic Berry phases
as a tool to calculate electronic Lorentz forces acting on the ions
in a general insulating system. The BO approximation, which amounts to
assume fixed and infinite mass nuclei, is made before the separation of
the center of mass motion. By these assumptions we neglect the motional
Stark effect and the mass correction terms.  The former is well justified
as long as the particle speed is small compared to the speed of light,
as in most condensed systems~\cite{note1}.  Mass correction terms become
important for exceedingly strong magnetic fields, when the insulating
character may be lost and/or particles may behave relativistically,
in which case these corrections should of course be included in the
nuclear and electronic hamiltonian.

We assume the ground state electronic wavefunction to be given,
at each time step, by an electronic structure calculation, such
as Hartree-Fock~\cite{ruder94}, or density functional theory (DFT)
methods~\cite{vignale}. The nuclear hamiltonian of a generic insulating
system in the BO approximation is
\beq
  \HcalN = \sum_n \frac{1}{2 M_n}
    \left[\Pvec_n - \frac{Q_n}{c} \Avec(\Rvec_n) - \chivec_n\right]^2 +
    U(\{\Rvec\}) \label{eq:nucl}.
\eeq
The first term is the ion kinetic energy ($\Pvec_n = -i\nabla_{\Rvec_n}$);
the second is the ground state expectation value of the electronic
hamiltonian, including the magnetic field effect on the electronic
structure. $\chivec_n$ appearing in the first term of eq.~(\ref{eq:nucl})
is the so called Geometric Vector Potential (GVP) or Berry connection,
given by~\cite{yin94,resta00}
\beq
  \chivec_n = -i \braket{\psiel(\{\Rvec\}) | \nabla_{\Rvec_n} |
    \psiel(\{\Rvec\})} \label{eq:gvp}
\eeq
where $\psiel(\{\Rvec\})$ is, in full generality, the many-body
electronic wavefunction in presence of the field, normalized to the
number of electrons: $\braket{\psiel|\psiel}= N_\el$, and depending
parametrically on all nuclear coordinate.  Whenever this wavefunction
can be chosen real and single-valued, the GVP of eq.~(\ref{eq:gvp})
vanishes.~\cite{resta00} In a magnetic field, the wavefunction cannot be made
real, and the GVP is nonzero. After integration of the electronic
degrees of freedom, a gauge potential associated with the GVP appears
in the nuclear hamiltonian~\cite{gaugefield,yin94,resta00}.  This gauge
field plays the role of an additional magnetic field, one that couples
only to the kinematic degrees of freedom, and not e.g., to the nuclear
moments, which instead experience the real field.  We restrict here for
simplicity to large gap insulators, where the adiabatic approximation
is well justified, and one can safely ignore all excited electronic
states in the expression of $\chivec_n$.  In that case moreover the spin
susceptibility is minuscule, and spin effects can also be neglected.

We derive from eq.~(\ref{eq:nucl}) the nuclear equations of motion
(EOMs) from the Heisenberg time evolution of the positions and velocity
operators.  Classical equations of motion are then obtained through
Eherenfest's theorem
\beq
  M_n \dot{\Vvec_n} = -\nabla_{\Rvec_n}U
    + (Q_n/c) \Vvec_n \times \Bvec
    + \sum_m \Vvec_m \times \Omegavec^{(nm)} \label{eq:eom}
\eeq
Eq.~(\ref{eq:eom}) resembles that of a charged ion in a field, but for
the last term, which is precisely the electronic Lorentz force. The
gauge invariant quantity $\Omegavec^{(nm)}$ is the Berry curvature which
plays the role of an effective magnetic field (gauge field) in the $3N-$
dimensional space spanned by the ionic degrees of freedom ($N$ is the
number of nuclei). It is given by
\beq
  \Omegavec^{(nm)} = -2\ \mathrm{Im} \Braket{\nabla_{\Rvec_n}\psiel |
    \times |\nabla_{\Rvec_m}\psiel} \label{eq:omega}
\eeq
and has the dimensions of a magnetic field, in fact proportional to the
external field $(B/c)$ when sufficiently weak.  Unsurprisingly, the force
on ion $n$ now depends upon the velocity of all other ions $m$ through
the off-diagonal terms in eq.~(\ref{eq:omega}).  If the electrons were
infinitely tightly bound to the nuclei (or equivalently, in the limit
of large separation between the ions), these off-diagonal terms would
vanish. In that regime, $\psiel$ collapses to a sum of products of single
particle orbitals centered around the nuclei. Each is rapidly decaying
in space, and each is dragged rigidly along by its nucleus as it moves,
thus providing total magnetic screening for every ion. Reality is of
course very far from that limit, magnetic screening is only partial,
and must be calculated explicitly.

The basic ingredient for computing the Berry curvature
eq.~(\ref{eq:omega}) is the derivative of the electronic wavefunction
with respect to atom position $\Rvec_n$. That can be obtained in an
electronic structure calculation by finite differencing the electronic
wavefunction $\psiel$ for two atomic configurations, compensating the
arbitrary phase of the wavefunctions (covariant derivative). Alternatively,
$\ket{\nabla_{\Rvec_n}\psiel}$ can be obtained by linear response to an
atom displacement.

In order to provide a first exemplification of the electronic Lorentz
forces, with a direct analytical and quantitative insight into the
properties of the Berry curvature eq.~(\ref{eq:omega}), we consider
here as a  simple example the classical dynamics a neutral homonuclear
diatomic molecule in a field.  If we set the molecule in motion and
freeze the vibrational and rotational degrees of freedom ($\Vvec_1 =
\Vvec_2 \equiv \Vvec$), the EOMs are
\beq
  M \dot{\Vvec} = (Q/c) \Vvec \times \Bvec
    +\Vvec \times \left(\Omegavec^{(11)}+\Omegavec^{(12)}\right)
    \label{eq:center}
\eeq
(inversion symmetry requires $\Omegavec^{(11)}=\Omegavec^{(22)}$ and
$\Omegavec^{(12)}=\Omegavec^{(21)}$). Since the molecule is neutral, the
total Lorentz force must vanish and $\Omegavec^{(11)}+\Omegavec^{(12)}
= -Q (\Bvec/c)$ must screen completely the nuclear charge $Q$. In the
heteronuclear diatomic case, the nuclei are not screened individually,
but only globally, in the form  $\sum_{nm} \Omegavec^{(nm)} = -(\Bvec/c)
\sum_n Q_n$~\cite{note2}.
If we consider a rigid rotation of the frozen
molecule, then $\Vvec_1 = -\Vvec_2 \equiv \Vvec$ around  the center of
mass and the EOM is
\beq
  M \dot{\Vvec} = (Q/c) \Vvec \times \Bvec
    +\Vvec \times \left(\Omegavec^{(11)}-\Omegavec^{(12)}\right)
  \label{eq:rotation}
\eeq
The electronic screening field $\Omegavec_\mathrm{r} \equiv
\Omegavec^{(11)} -\Omegavec^{(12)}$ depends upon the internuclear distance
and is related to the electronic Berry phase $\gamma$ accumulated on a
single period of rotation:
\beq
  \gamma \equiv \oint_{\partial C(R)} \chivec(R)\cdot d\Rvec =
  \int \!\!\! \int_{C(R)} \Omegavec_\mathrm{r}(R)\cdot\hat{\bm{n}}\,dS
  \label{eq:berryphase}
\eeq
where $C(R)$ is a circle of diameter $R$, $R$ being the interatomic
distance. As shown in Ref.~\onlinecite{ceresoli02}, $\gamma$ is the
electronic contribution to the rotation-induced magnetic moment.

We now wish to address the generically rotating, translating and vibrating
molecule. In general, that calculation can be done by first principle
implementation of eq.~(\ref{eq:eom}) and (\ref{eq:omega}). In order to
make the illustration more explicit while keeping it simple, we will focus
on a hydrogen molecule, where we can obtain essentially analytical results
by describing the electronic structure through Slater's variational LCAO
approximation based on two $1s$ orbitals~\cite{slater77}. To include
the field, we form linear combination of gauge-including atomic orbitals
(GIAO)
\bea
  \psiel &=& c_1\, \varphitilde_1(\rvec-\Rvec_1) +
             c_2\, \varphitilde_2(\rvec-\Rvec_2) \nn
  \varphitilde_n &=& \exp\left[-(i e)/c\,\Phi(\Rvec_n\rightarrow\rvec)\right]
  \varphi_{1s}(\rvec-\Rvec_n)
  \label{eq:giao}
\eea
where the phase factor $\Phi(\rvec\rightarrow\rvec')$ is the integral
of the vector potential along the line connecting the points $\rvec$
and $\rvec'$~\cite{TB}. $\varphi_{1s}$ is the hydrogenic $1s$ radial
wavefunction for a nuclear charge $\alpha$, taken as variational
parameter~\cite{slater77}: $\varphi_{1s}(\rvec) = (\alpha^3/\pi)^{1/2}\,
e^{-\alpha r}$.  For a fixed value of the internuclear separation,
$\alpha$ minimizes the sum of the electronic energy plus the
nuclear-nuclear coulomb repulsion, and varies from 1 in the limit of large
separation to $\sim$~2 in the limit of small separation where the H$_2$
molecule collapses to a He atom.  We neglect the electron spin due to the
large gap between the singlet and triplet ($\sim$~10~eV).  Owing to its
simplicity, this trial wavefunction gives for $H_2$ a crude equilibrium
distance of $1.00$~\AA\ (experimental $0.74$~\AA) and a barely reasonable
dissociation energy of 4.235~eV (experimental 4.476~eV).  However, it
illustrates the electronic Lorentz forces very well.  At weak field,
the one electron matrix elements are
\bea
  \widetilde{H}_{nm} &=& H_{nm}
    \exp\left[-(i e/c)\, \Phi(\Rvec_m\rightarrow\Rvec_n)\right] \nn
  \widetilde{S}_{nm} &=& S_{nm}
    \exp\left[-(i e/c)\, \Phi(\Rvec_m\rightarrow\Rvec_n)\right]
\eea
where $H_{nm}=\Braket{\varphi_n|H^0|\varphi_m}$ and
$S_{nm}=\Braket{\varphi_n|\varphi_m}$ are the field-free matrix elements.

\begin{figure}\begin{center}
  \includegraphics[width=\columnwidth]{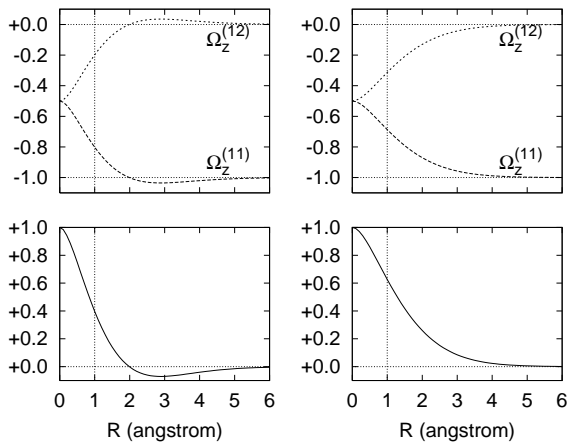}
  \caption{Top: $\Omega_z^{(11)}$ (solid line) and $\Omega_z^{(12)}$ 
  (dashed line) in units of $(B/c)$ as a function of the interatomic distance, 
  for the H$_2$ molecule. Bottom: The effective magnetic field felt by each 
  proton $1+\Omegavec_\mathrm{r}$, in units of $(B/c)$. Left panels, field
  normal to the molecule axis; right panels field parallel molecular axis.
  The vertical line indicates the theoretical equilibrium distance.}
  \label{fig:omega}
\end{center}\end{figure}

We computed  $\Omegavec^{(11)}$ and $\Omegavec^{(12)}$ with the trial
wavefunction of eq.~(\ref{eq:giao}), directly from eq.~(\ref{eq:omega}).
The resulting expression, analytic, contains a number of terms and is
too long to be shown here. Since GIAOs give the first order correction in
$(B/c)$ to the zero field wavefunctions, we retain only the linear $(B/c)$
term in the $\Omegavec$'s, where the zero order term is zero. These
quantities depend on the interatomic distance and on the angle between
the molecule axis and the magnetic field, preserving the cylindrical
symmetry of the system, independent of  the gauge. In the following
$\Omegavec^{(nm)}$ will indicate the coefficient of the first order term
in $(B/c)$.

The upper panels of Fig.~\ref{fig:omega} show $\Omegavec^{(11)}$
and $\Omegavec^{(12)}$ as a function of the interatomic distance,
for two relative orientations of the molecule with respect to the
magnetic field, which we choose parallel to the $z$ axis.  For large
interatomic separation $R$, i.e. in the dissociation limit, the
off-diagonal $\Omegavec^{(12)}$ vanishes and $\Omegavec^{(11)}$ reaches
the asymptotic value $-1$. This reflects the fact that the motion of the
two protons is decoupled in the dissociation limit and the electronic
screening of the individual nuclei is complete. For small interatomic
separation, both $\Omegavec^{(11)}$ and $\Omegavec^{(12)}$ (and by
symmetry $\Omegavec^{(22)}$ and $\Omegavec^{(21)}$) tend to $-1/2$,
recovering the correct screening of an isolated He atom. For arbitrary
interatomic distance, the sum of $\Omegavec^{(11)}$ and $\Omegavec^{(12)}$
is identically $-1$, which fulfills eq.~(\ref{eq:center}) and warrants
total screening of the center of mass motion. The lower panels of
Fig.~\ref{fig:omega} show the effective relative magnetic field at the
proton site, $(1+\Omegavec_\mathrm{r})$ as a function of the interatomic
distance, for two orientations of the molecule relative to the field. At
the equilibrium distance of $R\simeq 1$~\AA, the proton's Lorentz force
is that of a reduced  effective charge ranging between $+$0.4$|e|$  and
$+$0.6$|e|$, depending on orientation. (The true reduction is actually
weaker, since our variational wavefunction slightly overestimates the
electronic screening).  The approximate rotational g-factor~\cite{ramsey}
computed by eq.~(\ref{eq:berryphase}) in the present approximation
is 0.62, in fair agreement with experiment~\cite{ramsey}, and with
more accurate calculations~\cite{ceresoli02}, yielding 0.88. This sort
of error does not impair the value of the present approximation as an
analytical illustration of the method.

\begin{figure}\begin{center}
  \includegraphics[width=8cm]{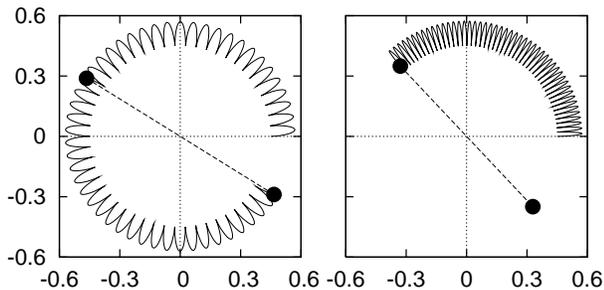}
  \caption{Simulated trajectories of the H$_2$, in presence of magnetic field.
  Left panel: neglecting the Berry curvature term. Right panel: with the
  Berry curvature term. In both cases, the molecule was set initially parallel
  to the $x$ axis and the bond compressed 0.1 \AA. The molecule rotates in the
  counter-clockwise sense. For a field 10 T, the estimated rotation period is
  of 2$\cdot$10$^{-5}$s. The units of the plots are \AA.}
  \label{fig:traj}
\end{center}\end{figure}

Armed with the Berry curvatures $\Omegavec^{(nm)}$ -- now known
analytically -- and with a simple parametrization of the interatomic
potential $U(R)$, we can describe the classical dynamics of H$_2$
in a field. As in a first principles molecular dynamics simulation,
we integrate the equation of motion eq.~(\ref{eq:eom}) for the H$_2$
molecule, exploring the effect of different initial (Cauchy) conditions%
~\cite{note3}. By way of example, we start by compressing or stretching
the molecular bond and let it free to vibrate/rotate. The restoring
force sets the two protons in motion initially in the radial direction
but soon their trajectory is deflected by the Lorentz forces. If the
molecule initially lies in the plane perpendicular to the magnetic field,
the resulting orbits resemble cycloids in that plane. The resulting
trajectories are shown in the upper panels of Fig.~\ref{fig:traj}. The
sense of rotation is determined by the initial condition of stretching
or compression.

To clarify the effect of the electronic Lorentz forces on the EOM we
show in the lower panels of Fig.~\ref{fig:traj} the same trajectories
now obtained by neglecting the Berry curvatures from eq.~(\ref{eq:eom})
-- i.e. only retaining the bare Lorentz force $(Q/c)\Vvec\times\Bvec$.
When the Berry curvature is included, the angular velocity of the
cycloid is reduced by the screening action of the electrons by a
factor $\sim$0.4, which is also the average fraction of magnetic field
felt by the ions during the vibrations around the equilibrium position
(see $\Omegavec_\mathrm{r}$ in Fig.~\ref{fig:omega}). We note that this
reduction factor measures the strength of the effective magnetic field
at the proton site, and differs from the rotational g-factor (here
0.62), which measures instead its integral over the orbit spanned in
a full rotation. Summing up, implementation of eq.~(\ref{eq:eom}) and
(\ref{eq:omega}) yields a description of nuclear motion whose accuracy
is only limited by that of the underlying electronic calculation.

All of the above is classical nuclear motion. In the quantum EOMs for the
nuclei, the presence of the GVP of Eq.~(\ref{eq:gvp}) must be considered.
In particular, starting from the zero field nuclear hamiltonian of a
diatomic molecule, we may explore how the GVP term of Eq.~(\ref{eq:gvp})
influences the roto-vibrational spectrum.  In the symmetric gauge,
$\Avec(\bm{x})=(1/2)\Bvec\times\bm{x}$ one can separate center of mass and
relative distance in Eq.~(\ref{eq:nucl}). The distance hamiltonian is
$\Hcal_\mathrm{R} = 1/(2\mu)\left[\pvec_R + (e/c)\Avec(\Rvec) + 
\chivec_R\right]^2 + U(R)$
where $\chivec_R = (e/2c)\Bvec\times\Rvec(S-1)/(S+1)$, $S$ being the
overlap integral between the two atomic orbitals. The curl of $\chivec_R$
is $\Omegavec_\mathrm{r}$. For ordinary laboratory magnetic field
intensities, the field can be considered as a perturbation, therefore we
expand the hamiltionian up to first order in the field
$\Hcal = \Hcal_0 + \Hcal_1$. $\Hcal_0$ is the unperturbed hamiltonian of a
harmonic vibrating rotator, $\Hcal_1 = (e B)/(2 \mu c) L_z 2S/(S+1)$ is the
rotational paramagnetic term.
The basis of the unperturbed hamiltonian $\Hcal_0$ can be labeled
by the quantum numbers $(n,l,m)$, and the spectrum is given by the
rotovibrational levels of the diatomic molecule.  
To first order in $(B/c)$, $\Hcal_1$ removes the
degeneracy of the rotational levels is removed according to the usual
Zeeman splitting $E_{nlm}=E^{(0)}_{nlm}+g_R \mu_{\rm{nuc}}\,B\,m$, where
$g_R$ is the rotational $g$-factor.  The eigenstates have corrections of
order $(B/c)$ due to the term $S/(S+1)$, which mixes an unperturbed state
with other states with the same $(l,m)$ and different $n$. However, for a
field of 1~T, the coupling between vibrational states is of the order
of 10$^{-5}$ cm$^{-1}$, much smaller than the centrifugal and anharmonic
couplings, which are of the order of 10~cm$^{-1}$.
To second order in $(B/c)$, there will be diamagnetic shifts affecting mainly
the states with $L_z=0$.
Despite the rotation, the canonical angular momentum $L_z$ is a conserved
quantity. In the symmetric gauge, the mechanical angular momentum is
\beq
  I \omega = L_z + \frac{e B}{2 c} \frac{2 S}{S+1} R^2 \label{eq:conserved}
\eeq
where $I$ and $\omega$ are respectively the momentum of inertia and
the angular velocity. The second term in the right-hand side is always
positive, depends on the vibrational state of the molecule, and is of
the order of $(B/c)$. The physical rotation of the molecule correspond to
the expectation value of $I\omega$; when $m\neq0$ the difference between
$I\omega$ and $L_z$ is negligible because it is of order $(B/c)$. However,
eq.~(\ref{eq:conserved}) shows that even when $L_z=0$, a small amount
of rotation still exists (see Fig.~\ref{fig:traj}).  

In conclusion, we presented a convenient formalism to calculate the
all important electronic contribution to adiabatic Lorentz forces for
atomistic dynamics in a magnetic field, based on Berry connections and
ideally suitable for future ab-initio simulations. We demonstrated
its validity and applicability in the simple example of H$_2$ where
variational wavefunctions provide approximate but analytical results for
the the Berry connection and Berry curvatures. The example demonstrates a
weak field induced coupling between rotations and vibrations. This work
is supported by MIUR COFIN No. 2004023199-003, FIRB RBAU017S8R operated
by INFM and MIUR FIRB RBAU017S8.



\begin{thebibliography}{10}
\bibitem{note1}
   The electric field experienced by a molecule moving in a uniform magnetic
   field is $\bm{\mathcal{E}}=\Vvec_\mathrm{CM}\times\Bvec$, where
   $\Vvec_\mathrm{CM}$ is the center of mass velocity. For gaseous hydrogen
   at 1000~K, the mean quadratic speed is $\sim$~4500~m/s. Thus
   $\mathcal{E}=$~4.5~$\cdot$~10$^{-7}$~V/\AA\ for a magnetic field of 1~T.
\bibitem{note2}
   There is also a radial component of Berry the curvature (in the direction
   of $\hat{\Rvec}$ in cylindrical coordinate). For this component the sum
   rule $\Omegavec^{(11)}+\Omegavec^{(12)}=-Q(\Bvec/c)$ implies that
   $\Omega_\mathrm{R}^{11}=-\Omega_\mathrm{R}^{12}$, since $\Bvec$, being
   uniform, has no radial component.
\bibitem{note3}
   We integrated the EOMs by an adaptive time step 4th-order Runge-Kutta
   scheme. Velocity Verlet and predictor-corrector lead to an irreversible
   energy transfer between vibrational and rotational degrees of freedom.
\end{thebibliography}
\end{document}